\documentstyle[prb,aps]{revtex}

\begin{document}

\centerline{\bf {A New Hypothesis for Layers of High Reflectivity Seen in 
MST Radar Observations }}

\vspace{1cm}
\centerline{\large G. C. Asnani$^\star$ and M. K. Rama Varma Raja$^\star \star$}

\vspace{0.3cm}
\centerline{$^\star$ Indian Institute of Tropical Meteorology, Dr. Homi Bhabha Road
, Pune 411 008(India). E-mail: asnani@giaspn01.vsnl.net.in or Prof.asnani@vsnl.com}
\centerline {$^\star \star$ Department of Physics, University of Pune, Pune 411 007(India). E-mail: rama@physics.unipune.ernet.in}

\vspace{1cm}

We have worked with MST Radar located at Gadanki, Tirupati, India(13.47 degree N, 
79.18 degree E). Altitude of the location is approximately 100 meters. This is the
first and only MST Radar operating in India. It is in Tropical Monsoon region. Monsoon 
moves northward across Tirupati around 1st June and withdraws southward across
Tirupati in the first week of December. Our interest has been to examine the
characteristics of vertical velocity and other wind characteristics associated
with monsoon. We examined the characteristics of range-compensated Signal-to-Noise
ratio ($r^2$ SNR). We call this as Reflectivity. We examined the MST Radar
data set for 14 months (September 1995 to November 1996 except May 1996). The height
range considered for the analysis was between 3.6 km and 21 km above ground. Our
findings have been on the following lines:

1) Except on Thunderstorm occasions, high reflectivity-regions are in the form
of layers 1-2 km thick. One such high-reflectivity layer is always present near
tropopause level ~ 17 km. Below the tropopause, there is a bunch of such 
high-reflectivity layers, generally 3-4 in number, between 4 and 11 km. The 
atmospheric layer between 11 and 15 km is generally free from high reflectivity 
layer.

2) On a thunderstorm occasion, there is a deep high-reflectivity layer extending
from 4-11 km. After the passage of the thunderstorm, this deep layer of high
reflectivity breaks up into layers.

3) Layers of high reflectivity occur throughout the year even outside monsoon 
season, when ITCZ is far away from the region and they cannot be attributed
as originating from deep convective clouds. When visual observation and MST
observations are taken simultaneously at Gadanki, visual clouds are estimated
to be at the same levels as the high reflectivity layers seen through MST
observations. Indian satellite pictures over Gadanki also suggest similar
heights of clouds as given by MST Radar for high reflectivity layers.

4) Satellite pictures in infra-red range show much more extensive areas of
cloudiness than the pictures in the satellite visible range. In other words,
there are extensive sub-visual clouds in the atmosphere.

5) Every month early morning and late evening, in twilight hours we see beautiful
cirrus clouds in the form of cloud streets, streaks or sheets on several
occasions which may or may not be clearly visible at other hours.

6) On looking at the sky frequently, one gets the impression that the sky is not
clear even though we do not see clouds or haze layers. In these haze layers, appear
clouds in the form of cloud streets, streaks or cloud sheets at some times.
The appearance of these clouds gives the clear impression that there are 
waves in the atmosphere which give visible clouds in the region of upward 
motion associated with these waves. Again in each bigger wave cloud, there are
smaller and smaller wave clouds and clearances; there are waves within waves.
When the clouds disappear, they leave a sort of haze. Hence, the clouds form out 
of haze and leave some haze after dissipating. Even when there are no visible
clouds during the day or night, the structure of the atmosphere is patchy in
appearance. One gets a clear impression that there are waves and waves, bigger
and smaller waves in the atmosphere which are creating patches of haze and
sometimes visible clouds in the atmosphere. These are due to gravity-type
waves with a very wide spectrum of horizontal and vertical wavelengths. As we
know from theoretical and observational evidence, horizontal wavelengths are
1-2 orders of magnitude larger than vertical wavelengths. If and when they
occur in the atmosphere, they have a tendency to take horizontally spread
layered structure with vertical depth 1-2 orders of magnitude smaller than
horizontal extent.

7) In the atmosphere, we visualize three classes of waves:
a.Inertial waves or Rossby-type waves: Their horizontal extent is of the order
of a few thousand kilometers and vertical wavelengths of the order of 10 km. In
the mechanism of their formation, we have to consider the rotation of the earth
and the resulting coriolis force. Their period is of the order of a few days.
b. Gravity waves: These arise mainly from local horizontal pressure gradients
arising out of gravitational weight of the overlying air column, air accelerating
from higher pressure towards lower pressure. Horizontal accelerations and displacements
are accompanied by appropriate vertical accelerations and displacements to
conform to the requirements of law of conservation of mass i.e., equation of
continuity.

  Vertical displacements and accelerations of air parcels can also arise from
buoyancy forces. Heavier parcel tends to sink down while lighter air parcels
tend to rise up in an environment of horizontal density gradients. These are
called Brunt-Vaisala oscillations. These horizontal gradients of density arise
out of differences in temperature, humidity and hydro-meteor loading.

 This hydrometeor-loading needs a little elaboration:

 (i) Every parcel of
cloud air contains at least one hydrometeor in liquid water or solid form.
Inside each hydrometeor is an aerosol which acts as a nucleus which has induced
condensation and/or freezing.

(ii) Invariably, hydrometeor has higher density
than the surrounding air. As such, it tends to fall down due to gravitational
force.

(iii) As it descends down, it exchanges sensible heat and moisture
with its environment. Hence, its volume, mass, and density undergo a change
during its descent.

(iv) As the hydrometeor descends down with gravitational acceleration through the
air parcel, there is frictional resistance/viscous resistance to its vertical
motion. soon, the hydrometeor loses its acceleration and descends down with
what is called "Terminal velocity". Where has its weight gone? Its weight is
taken up by the air parcel which is offering resistance to its vertical 
movement and acceleration. In other words, the air parcel becomes heavier
to the extent it has taken over the gravitational acceleration of the falling
hydrometeor. If the hydrometeor falls with its terminal velocity, it means that its
entire weight is taken over by the air parcel. As such the air parcel has become
heavier by the total weight of the hydrometeor; the density of the air parcel
has effectively increased.

If only half of the gravitational acceleration of the descending hydrometeor has been
taken over by the air parcel, then the weight of the air parcel has increased
only by half the weight of the hydrometeor.

The hydrometeor gradually loses its gravitational acceleration and gives its
weight to the surrounding air parcel gradually and not instantaneously. As such,
the air parcel takes the load of the hydrometeor gradually during the downward
trajectory of the descending hydrometeor.
 
In many calculations, for the sake of simplicity, it is assumed that as soon
as condensation or freezing takes place in the atmosphere, the load of the
hydrometeors is immediately taken over by the surrounding air. However, we have
to recognize that the hydrometeor-loading occurs gradually through a finite
interval of time.

We should also remember that during the fall, the hydrometeor is simultaneously
undergoing a change in its volume, mass and density. Hence, a realistic, quantitative
estimate of hydrometeor loading effect on the air parcel needs careful calculation.
However, nature takes care of the process and creates varying density effects on
the cloud air parcel as the hydrometeor descends.

(v) In addition to buoyancy fall of the hydrometeor, there are upward and downward
motions of air parcels inside the cloud. These upward and downward motions of
air parcels inside the cloud create further complications in the calculation of
hydrometeor-loading effect on air parcel.

(vi) In addition to pure buoyancy forces operating in a class of waves called
gravity waves, there also occur what are known as Kelvin-Helmholtz waves
due to presence of vertical shear of horizontal winds which is almost
always present.

(vii) This class of gravity waves of Kelvin-Helmholtz type have very small wavelengths
of the order of centimeters and meters and correspondingly small periods of the
order of a few minutes. Earth's rotation or Coriolis force does not perceptibly
come into the calculations for these waves.

c. Inertio-Gravity waves: Between the two extremes of large inertial waves and small
gravity waves, there is an intermediate class of waves which may be termed as
inertio-gravity waves in which Coriolis force plays some role along with gravity
force. There is literature on the subject of inertio-gravity waves (
\cite{mur},  \cite{wur}, \cite{jos}, etc.), but more work needs
to be done on this class of waves. Orographic influence also come into play.

8) When we fly in an aircraft, large-scale weather and clouds are influenced by
Rossby-type waves. When we look at the sky through cockpit or through the
window near the window-seat, during day time, we immediately get the view of
air clouds at different levels and also gravity waves of various dimensions near the
flight level. We see fairly large waves with estimated wavelengths of the order
of tens of kilometers, along with embedded smaller and smaller waves, thick clouds,
thin clouds, thinner clouds, space filled with haze and space clear of visible clouds.

In our view, MST Radar reflectivity pattern gives us spot view of these numerous beautiful
waves.

9) We have analyzed the field of MST Radar reflectivity as seen at Gadanki along
with MST Radar measured wind fields. We have interpreted the reflectivity 
fields within a conceptual model given below:

(i) Inertio-gravity waves in the atmosphere generate layers of upward/downward
motion, high/low humidity and high/low temperature lapse rates. The layers of
upward and downward motion are regularly seen in the vertical wind field given
by MST Radar. The vertical wavelength of these inertio-gravity waves has wide
spectrum depending on orography and diabatic heating; vertical wavelength of
about 5 km is a more frequently observed wavelength. The corresponding 
horizontal wavelength is of the order of 200 km, the more dominant wavelength
visible in satellite picture is 500 km in the direction of wind and 1000 km
across the wind.

  The layers of high relative humidity created by inertio-gravity waves are
favorable for the formation of layered clouds, which we call "Mother Cloud
Layers"; these clouds may be visible or sub-visible.

(ii) Hydrometeors inside a "Mother Cloud Layer" tend to fall down attaining
their respective terminal velocities. During their stay inside the clouds, the
hydrometeors exchange heat, moisture, mass and momentum with the environmental
air on small micro-scales. These exchanges between the hydrometeors and the
"Mother Clouds's air" create strong gradients of temperature, humidity, density
and momentum. Density variations are also created through hydrometeor-loading.

 Electrical charges are also generated during the processes of condensation,
evaporation, freezing, melting and sublimation.

 These micro-physical gradients in density along with prevailing wind field in
the vertical generate internal gravity waves in the form of Brunt-Vaisala 
oscillations, Kelvin-Helmholtz waves and other waves of different horizontal
and vertical wavelengths. These wavelengths range from a few millimeters to
tenths of meters in the vertical and from a few meters to about thousand
meters in the horizontal. Known laws of physics and dynamics suggest that the
wavelengths may be still smaller, equivalent to the distances between adjacent
parcels of air exchanging heat moisture, mass and momentum, with the 
hydrometeor embedded in the parcel.

(iii) The strong gradients of temperature, humidity and density created by 
these micro-physical and micro-dynamical processes in the air surrounding the
hydrometeor or an ensemble of hydrometeors cause strong variations in the
refractive index of air parcel, in respect of electro-magnetic lidar and
radar beams impinging on the air parcels. In turn, this causes high 
reflectivity/scatter of the impinging lidar/radar beam. In respect of VHF
MST radar Bragg-type reflection/scatter is a dominant type of reflection/
scatter.
     
     We examined the size of air parcels giving the highest values of 
reflectivity, at Gadanki. We came to the conclusion that their horizontal
extent is of the order of 1 km while their vertical extent is of the order
of 100 m.

 Indian MST radar beam oriented in vertical has a half-wavelength of about
3m. As such, Indian MST radar is capable of detecting reflectivity patterns
of vertical wavelengths of the order of about 6m. These small-scale 
variations in reflectivity appeared in the form of very delicate embroidery
inside the large-scale reflectivity of the "Mother Cloud Layer".

(iv) These variations in the reflectivity pattern may look like turbulent
fluctuation in the atmosphere. If we do not associate these fluctuations
with clouds and hydrometeors inside the clouds, the clouds might appear
as clear-air turbulence as has been prevalent in the current explanation
appearing in literature connected with MST radars. This prevalent explanation
has faced paradoxes, the main paradox being of thin horizontal sheets of
turbulence.

   If we free our thinking from the concept of clear-air turbulence and turn 
our thinking in the direction of visible or sub-visible clouds containing
hydrometeors, the apparent paradoxes of clear-air turbulence causing high-
MST radar reflectivity get immediately resolved.

   The existence of sub-visible clouds occupying much larger area than the
visible clouds has now been established beyond question, through latest
observations by satellites in the infra-red range, by aircraft flying
through cirrus cloud air and by lidars operating in high altitude aircraft
sending their beams through visible and sub-visible cloud layers.

(v) Using over 2,50,000 observational data points for MST radar reflectivity
and vertical wind shear (deduced from corresponding MST radar wind observations)
spread over 14 months (from 1995 September to 1996 November), we plotted 
scatter diagrams of MST radar reflectivity versus vertical wind shear. We are
pleasantly surprised to find that reflectivity decreases almost exponentially
as vertical wind shear increases. If mechanical turbulence was the main cause 
of high reflectivity, we should see reflectivity increasing with vertical
wind shear, and not decreasing almost exponentially. Scatter diagrams for
each of the 14-months are presented in \cite{ram1}. Also, Scatter diagrams 
for four representative months (January, April, July and October), 
for Indian monsoon region, are presented in \cite{ram2},\cite{ram3}.
This shows that mechanical turbulence is
not the principal cause of high-MST radar reflectivity. 
\cite{asn} had hypothesized that turbulence may not be the 
primary cause of high-reflectivity seen in MST radars.

(vi) Knowing the importance of cirrus clouds, the world scientific community
launched the programme known as FIRE (First ISCCP Regional Experiment). This
FIRE programme concentrated on the study of layer clouds (Cirrus Clouds in the
upper troposphere and low level stratus clouds in the lower troposphere).
FIRE I programme was executed during the period 1985-1990 while FIRE II
programme was executed during the period 1990-1995. FIRE III programme is 
proposed to be executed in the beginning of this new millennium, with 
particular emphasis on the tropics. Results of FIRE I have been summarized
in a special issue of Monthly Weather Review (November, 1990); results of
FIRE II have been summarized in a special issue of Journal of Atmospheric
Sciences (December, 1995). This topic is an important component of CLIVAR
programme in the tropics.

  The results of FIRE I and FIRE II have broadly confirmed that there is fine
micro structure inside the layered clouds which can be interpreted easily,
atleast qualitatively, in terms of micro-physical and micro-dynamical processes
presented above. 

 Also, the latest Numerical Modeling Work on Cirrus Clouds (for example
\cite{sas}, \cite{khv1},\cite{khv2}) show that there
are fine-scale structures of various in-cloud parameters including temperature,
humidity and ice-concentrations.

 In fact in our view these fine scale structures created inside the visible
 or sub-visible "Mother Cloud Layers" can cause steep refractive index 
gradients sufficient to cause high-MST radar reflectivity.

10) We expect that the conceptual model of MST radar reflectivity presented
above will give a new orientation to the thinking and interpretation of
MST radar reflectivity. It will provide a satisfactory, physically and 
dynamically acceptable, interpretation of the reflectivity patterns seen in
the MST radar observation.

11) A few corollaries follow from this conceptual model:

(i) Since, high-MST radar reflectivity layer, 1-2 km thick, is always observed
near the tropopause, it may or may not be directly connected to inertio-gravity
waves. The mechanism for the formation and sustenance of high-reflectivity 
layer is realized as follows:

(a) The temperature near the tropopause level are very cold (~200 K); 
temperature lapse rate is stable, 2-4 degree C/km; as such vertical mixing
of air is inhibited and is weak. Relative humidity below the tropopause is
high \cite{new}. Water substance and aerosols injected into the
upper troposphere by deep convection near ITCZ remains below the tropopause
while the relative humidity is very low above the tropopause. Water substance
and aerosols form visible or sub-visible cirrus cloud layer or haze layer below
the tropopause. Through the micro-physical and micro-dynamical processes
mentioned earlier, the air layer develops strong vertical gradients or
discontinuities in the refractive index with respect to MST radar beam
impinging on the air there. This gives high-reflectivity echoes near the tropopause level on all days of the year.

   When the cloud gets dissolved, it leaves large number of aerosols suspended 
there. By themselves, aerosols are not likely to be detected by MST radar.
But the layer of high content of aerosols can be detected and has been
detecting near the tropopause (\cite{nee1},\cite{nee},\cite{sch} etc.).

  In the tropics, ITCZ injects aerosols and moisture which remain trapped
in the layer immediately below the tropopause. In extra-tropics, this 
injection is done by extra-tropical cyclone waves and polar front. On
the same reasoning, as given for tropics, a layer with high content of aerosols
and thin cirrus ice crystals will also get formed below the tropopause
in extra-tropics. Thus a layer having high content of aerosols and thin cirrus
cloud is expected to envelope the whole earth's atmosphere near tropopause.
This has been substantiated by observations of (Nee et al., 1995, 1998; FIRE I
and FIRE II observations and their results published in the Special issues of
Monthly Weather in November 1990 and Journal of Atmospheric Sciences in
December 1995 respectively). Asnani et al. \cite{asn1} had 
hypothesized the existence of an aerosol layer near tropical tropopause, 
based on the same mechanism mentioned above.

(ii) As we have stated above, observations confirm existence of alternate
layers of upward and downward motion with vertical depth of 2-3 km. We
should expect accumulation and depletion of both water substance and aerosol
substance near the levels of vertical convergence. Further vertical upward
motion tends to create higher relative humidity and higher temperature lapse
rate; vertical downward motion tends to do the opposite, creating stable
lapse rate and drier air. Hence we should expect to find layers of high
and low relative humidity in the vertical. Sensitive instrumentation is required
to detect this type of structure in the atmosphere; this structure is likely
to be missed by the ordinary radiosonde instruments. Such layers have been
detected by special effort \cite{ise}.

(iii) As stated earlier, these alternate layers of upward and downward motion 
are associated with inertio-gravity waves, which are always present in the
atmosphere. These vertical wave motion will also tend to split large convective
clouds, particularly in their decaying stage, into layered clouds. While
convective instability in the atmosphere tends to generate deep convective
clouds in the tropical atmosphere, these inertio-gravity waves inhibit
the formation and sustenance of these deep convective clouds in the tropical
atmosphere.

(iv) Cirrus clouds have a capacity to retain their existence for a long time,
even away from the source of their formation. Hence, visible or sub-visible
cirrus clouds are likely to be seen at many places, with or without deep
convective clouds. The upward vertical motion associated with inertio-gravity
waves tends to generate cirrus clouds, visible or sub-visible at many places
in the atmosphere. If nothing else, these clouds influence the radiative
heat budget of the earth's atmosphere. 

Acknowledgment: The authors thank Prof. D. Narayana Rao and his group at
Department of Physics, Sri Venkateswara University, Tirupati (India) for
considerable help in Indian MST Radar data collection and analysis. 
Mr. M. K. Rama Varma Raja thanks Dr. (Mrs.) P. S. Salvekar and the Director, Indian Institute
of Tropical Meteorology for providing necessary support and facilities during 
the course of research work.


\begin{references}

\bibitem{mur} 
Muraoka, Y., T. Sugiyama and K. Kawahira, 1988: Cause of a monochromatic
inertia-gravity wave breaking observed by the MU radar, Geophys. Res. Lett.,
15, 1349-1352.

\bibitem{wur} 
Wurtele, M. G., Datta, A. and Sharman, R. D., 1996: The propagation of gravity-
inertia waves and Lee waves under a critical level. J. Atmos. Sci., 53,
1505-1523.

\bibitem{jos} 
Joseph, B., 1997: Chaotic mixing by internal inertia-gravity waves. Phys. 
Fluids, 9, 945-962.

\bibitem{ram1} 
Rama Varma Raja, M. K., 1999: " Parameterization of Cloudiness Over India and
Neighbourhood ", Ph.D. Thesis submitted to University of Pune, Pune-411007,
India.

\bibitem{ram2} 
Rama Varma Raja, M. K., G. C. Asnani and P. S. Salvekar, 1999a: MST Radar Data
can Reveal Micro-Structure of Clouds". Preprints of 29th Conference on
Radar Meteorology (American Meteorological Society), Montreal, Canada 
(July 1999), pages 343-346.

\bibitem{ram3} 
Rama Varma Raja, M. K., G. C. Asnani, P. S. Salvekar, A. R. Jain,
D. Narayana Rao, P. Kishore, S. Venkoba Rao and M. Hareesh, 1999b: Layered
Clouds In The Indian Monsoon Region. Proc. Ind. Acad. Sci. (Earth and Planetary
Sciences), Paper accepted for Publication in the December 1999 issue of the
Journal. 

\bibitem{asn} 
Asnani, G. C., M. K. Rama Varma Raja, D. Narayana Rao, P. S. Salvekar,
P. Kishore, T. Narayana Rao and P. B. Rao, 1998a: Patchy Layered Structure
of Tropical Troposphere As Seen By Indian MST Radar. STEP HANDBOOK (As a
Proceedings of the 8th International Workshop on Technical and Scientific
Aspects of MST Radar held at Bangalore during 15-20 December, 1997), Editor
Belva Edwards, SCOSTEP SECRETARIAT, BOULDER, CO, USA, pages 192-195.

\bibitem{sas} 
Sassen, K. and Dodd, G. C., 1989: Haze particle Nucleation Simulations in 
Cirrus clouds and Applications for Numerical Lidar Studies. J. Atmos. Sci., 46,
3005-3014.

\bibitem{khv1} 
Khvorostyanov, V. I. and Sassen, K., 1998a: Cirrus Cloud Simulation Using
Explicit Microphysics and Radiation. Part I: Model description. J. Atmos. Sci.,
55, 1808-1821.

\bibitem{khv2} 
Khvorostyanov, V. I. and Sassen, K., 1998b: Cirrus Cloud Simulation Using 
Explicit Microphysics and Radiation. Part II: Microphysics, Vapor and Ice mass
budgets, and optical and radiative properties. J. Atmos. Sci., 55, 1822-1845.

\bibitem{new} 
Newell, R. E., Zhu, Y., Reed, W.J., and Waters, J. W., 1997: Relationship 
between tropical upper tropospheric moisture and eastern tropical pacific
sea surface temperature at seasonal and interannual time scales. 
Geophys. Res. Lett., 24, 25-28.

\bibitem{nee1} 
Nee, J.B., G.B. Wang, P.C. Lee and S. B. Lin, 1995: Lidar studies of particles
and temperatures of the atmosphere: First results from National Central
University lidar. Radio Science, 30, 1167-1175.

\bibitem{nee} 
Nee J. B., Len C. N. and Chen, W. N., 1998: Lidar observations of the cirrus
cloud in the tropopause at Chung-Li (25 degree N, 121 degree E). J. Atmos. Sci.,
55,2249-2257.

\bibitem{sch} 
Schroder, F. and Strom, J., 1997: Aircraft measurements of sub-micrometer
aerosol particles (< 7nm) in the midlatitude free troposphere and tropopause
region. Atmos. Res.,44, 333-356. 

\bibitem{asn1} 
Asnani, G. C., M. K. Rama Varma Raja and P. S. Salvekar, 1998b: Aerosol Layer
Near Tropical Tropopause. J. Aerosol Sci. (USA), 29,suppl.1, pp. s649-s650.

\bibitem{ise} 
Iselin, J. P. and W. J. Gutowski Jr., 1997: Water Vapour Layers in Strom-Fest
Rawinsonde Observations, Mon. Wea. Rev., 125, 1954-1963.

\end{references}
\end{document}